\documentstyle[11pt,newpasp]{article}
 
\def\deg{$^{\circ}\,$}

\def\etal{{\it et al.\ }}

\def\ie{{\it i.e.\ }}

\begin{document}

\title{Centers of Barred Galaxies: Secondary Bars and Gas Flows}
\author{Witold Maciejewski} 
\affil{Max-Planck-Institut f\"ur Astronomie, Heidelberg, Germany}

\section{Introduction}
Active Galactic Nuclei (AGN) require mass accretion
onto the central engine. On large galactic scales, torques from a stellar bar 
can efficiently remove angular momentum from gas, and cause it to move inwards
along two hydrodynamical shocks on the leading edges of the bar. Inflowing gas
settles on near-circular orbits around the Inner Lindblad Resonance (ILR), and
forms a nuclear ring, about 1 kpc in size. A secondary bar inside the main 
one, with its own pair of shocks, has been proposed to drive further inflow,
and thus feed the AGN in a manner similar to the inflow on large scales. Here, 
we report results of high resolution hydrodynamical simulations, where we 
examine the nature of the nuclear ring, and check how efficient double bars 
can be in fueling AGNs.

\section{Hydrodynamical models with a single bar}
All calculations have been done with a grid-based PPM code in 2 dimensions, 
for isothermal, non-selfgravitating gas, and point symmetry has been imposed. 
Excellent resolution near the galaxy center (better than 20 pc inside the 
nuclear ring) was achieved by using a polar grid. In order to trace shocks 
better, we calculated the value div$^2{\bf v}$ throughout the grid, where 
div {\bf v} $<0$, and displayed it next to the density diagrams.

The nuclear ring at low gas sound speeds ($c_s=5$ km s$^{-1}$) is made of a 
tightly wound spiral (Piner, Stone \& Teuben 1995 ApJ 449, 508) with no 
shocks (Fig.1 {\it left}). The pair of straight main shocks in the 
bar ends clearly at the outer edge of the nuclear ring, and only a weak, 
tightly wound sound wave propagates through gas inside the ring. There is no 
significant gas inflow to the center.

The exceptionally high resolution of our method allowed us to study the 
structure of the gas flow at high sound speed (20 km s$^{-1}$)
Instead of the nuclear ring, a nuclear spiral with higher pitch angle 
develops (Fig.1 {\it center}). Its presence on 
the div$^2{\bf v}$ plot indicates a spiraling shock, along which the gas falls
towards the center. We predict a significant gas inflow here, which can fuel
an AGN.

\section{Results for a dynamically possible double bar}
In a dynamically possible double bar, the primary bar must have an ILR, and 
the secondary bar must end well within the outer ILR of the main bar
(Maciejewski \& Sparke 1999 MNRAS {\it in print}). Resonant coupling favors the
existence of stable orbits supporting double bars (Tagger \etal 1987 ApJ 318, 
L43). Coupling the corotation resonance of the secondary bar with the outer ILR
of the main bar causes the dynamically possible secondary bar to
end well inside its own corotation, \ie it rotates slowly. 

A snapshot of gas flow in a dynamically possible system with two independently
rotating bars (Fig.1 {\it right}) shows the nuclear ring widened or destroyed 
by the secondary bar. An elliptical ring develops around the inner bar, with 
a size that is largely independent of the sound speed: the flow is mainly 
elliptical with weak transient shocks. Straight shocks form in fast rotating 
bars only --- they curl around a slow bar and turn into a ring (Athanassoula
1992, MNRAS 259, 345). No significant gas inflow to the center is seen, even 
at high sound speed. 

\section{Conclusions}
In a singly barred galaxy both near-circular motion 
of gas in the nuclear ring, and a spiraling shock extending towards
the galaxy center are possible, depending on the sound speed in
the gas. A dynamically possible doubly barred galaxy is likely to have 
a slowly rotating secondary bar, which neither creates shocks in the 
gas flow, nor enhances gas inflow to the galaxy center.

\begin{figure*}[b]
\includegraphics{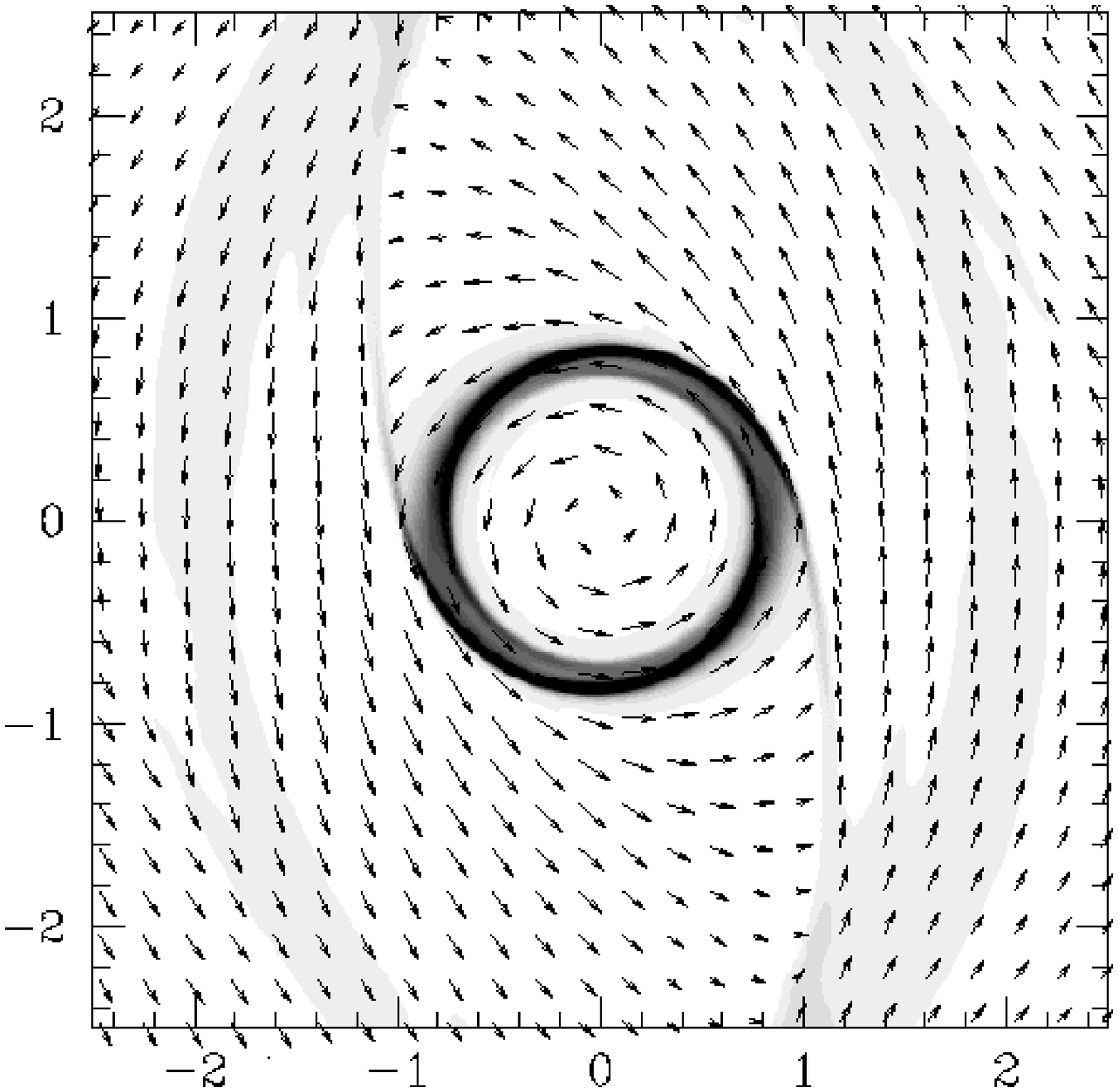}
\includegraphics{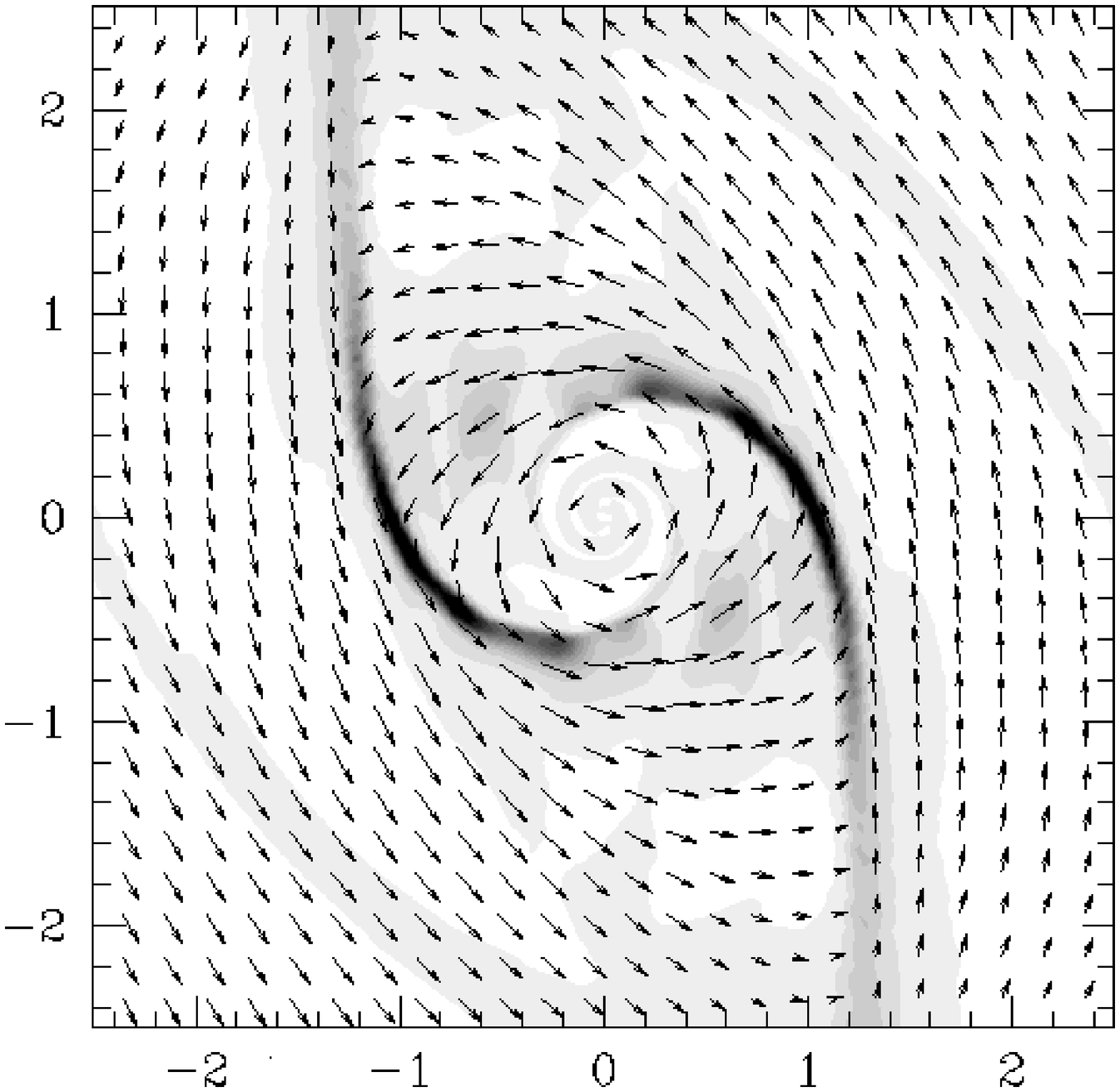}
\includegraphics{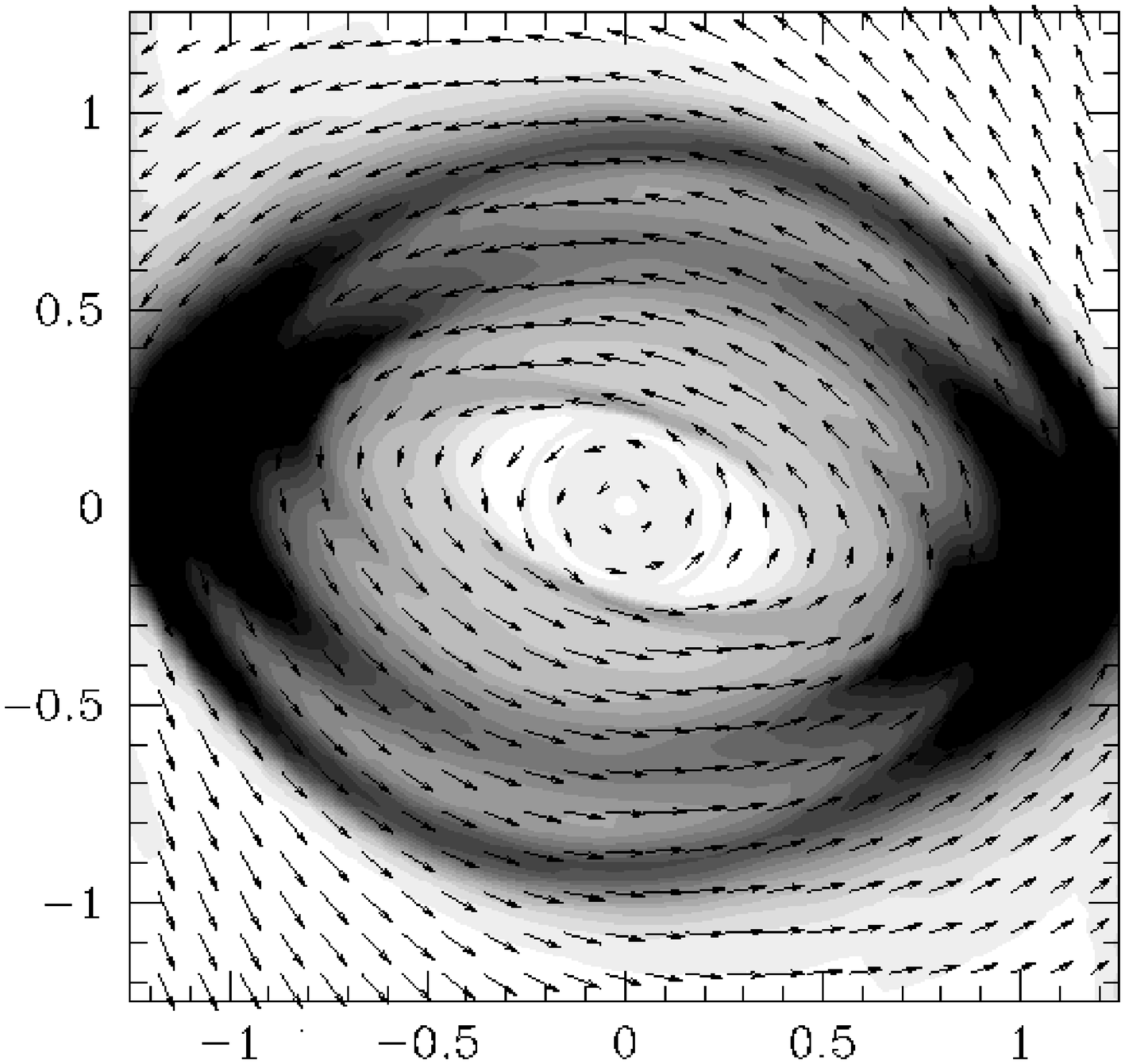}
\includegraphics{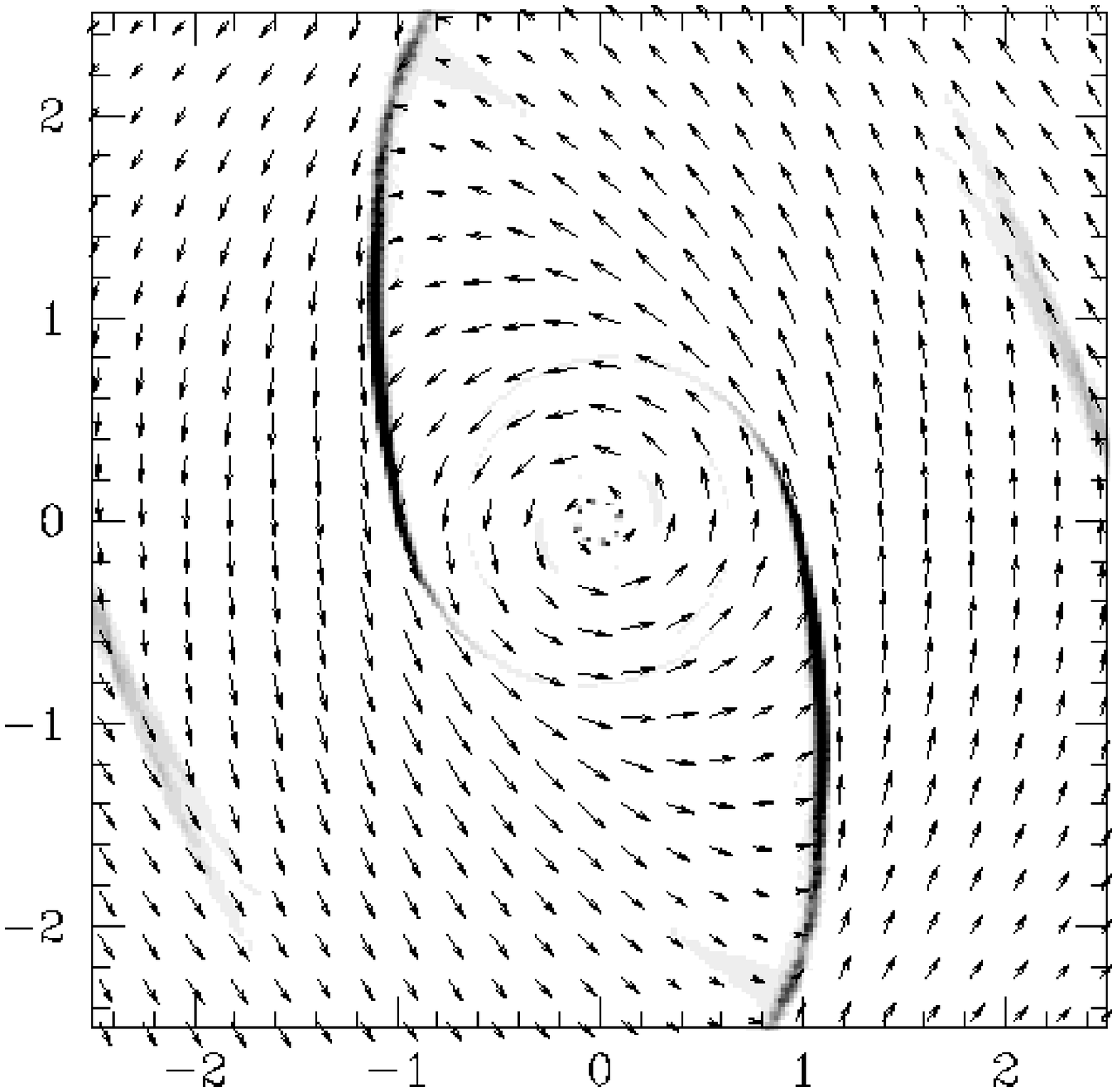}
\includegraphics{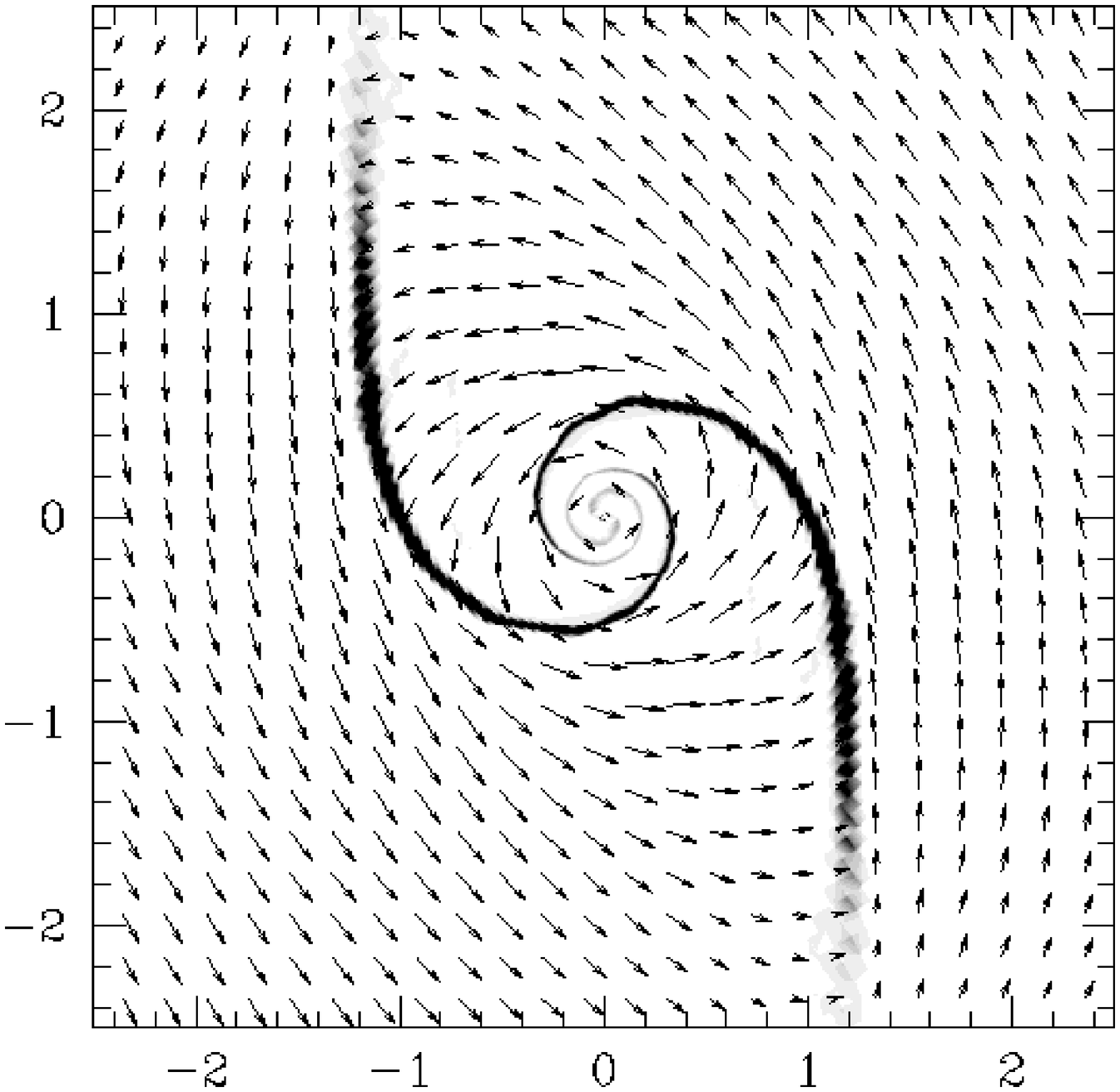}
\includegraphics{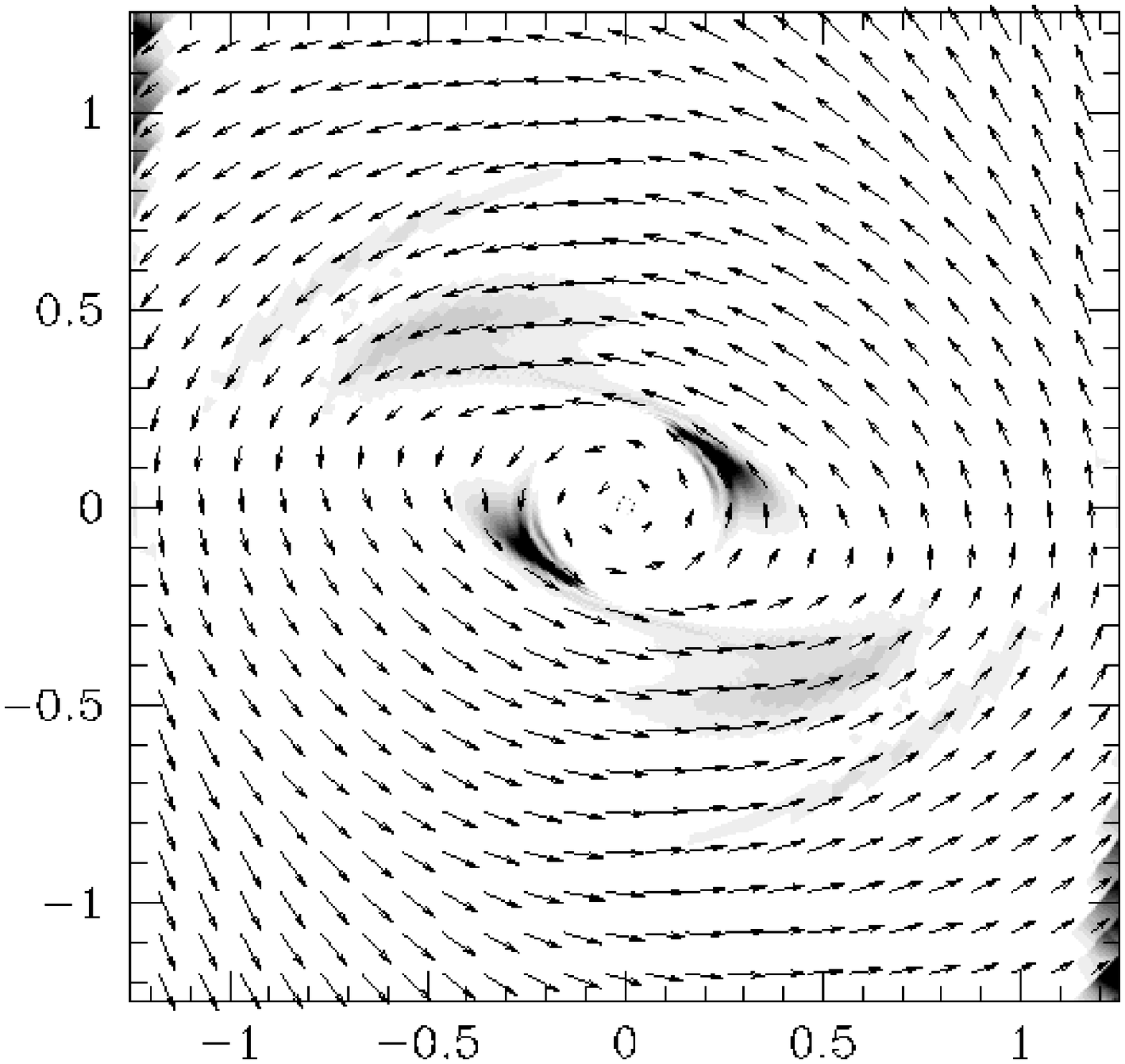}
\vspace{25mm}
\caption{Density ({\it top}) and div$^2{\bf v}$ ({\it bottom}) diagrams for 
inner regions of modeled gas flow in barred galaxies with ILRs. The velocity 
field is shown in the frame rotating with the main bar, which is vertical 
on all frames, rotates counterclockwise, with corotation resonance at 5.8, 
and outer ILR at 2.3 (units are in kpc). {\it 
Left:} flow in a single bar at $c_s=5$ km s$^{-1}$. {\it Center:} same for 
$c_s=$ 20 km s$^{-1}$. {\it Right:} flow in a double bar at 
$c_s=5$ km s$^{-1}$; the angle between the bars is 78\deg.}
\end{figure*} 
\end{document}